\documentclass[12pt]{article}
                          
\usepackage{slashed}
\usepackage{axodraw}
\usepackage{pix}

\usepackage{epsfig}
\usepackage{calrsfs,bbm,bm}
\usepackage{color,cancel,graphicx,mathrsfs,psfrag}
\usepackage{dsfont}
\usepackage{amsmath}
\usepackage{amssymb}
\usepackage{exscale}

\def\a0size{6}

\newcommand{\lsi}{\raise0.3ex\hbox{$<$\kern-0.75em\raise-1.1ex\hbox{$\sim$}}}
\newcommand{\gsi}{\raise0.3ex\hbox{$>$\kern-0.75em\raise-1.1ex\hbox{$\sim$}}}
\newcommand{\lsim}{\mathop{\lsi}}

\renewcommand{\vec}[1]{{\bf #1}}

\newcommand{\mbar}{\overline{m}}
\newcommand{\nF}{n _ {\rm f } }
\newcommand{\ord}{O}

\allowdisplaybreaks[1]

\DeclareMathOperator*{\SumInt}{%
\mathchoice%
  {\ooalign{$\displaystyle\sum$\cr\hidewidth$\displaystyle\int$\hidewidth\cr}}
  {\ooalign{\raisebox{.14\height}{\scalebox{.7}{$\textstyle\sum$}}\cr\hidewidth$\textstyle\int$\hidewidth\cr}}
  {\ooalign{\raisebox{.2\height}{\scalebox{.6}{$\scriptstyle\sum$}}\cr$\scriptstyle\int$\cr}}
  {\ooalign{\raisebox{.2\height}{\scalebox{.6}{$\scriptstyle\sum$}}\cr$\scriptstyle\int$\cr}}
}

\begin{document} 

\setlength{\baselineskip}{0.6cm}
\newcommand{\figysize}{16.0cm}
\newcommand{\figtopspace}{\vspace*{-1.5cm}}
\newcommand{\figbottomspace}{\vspace*{-5.0cm}}
  
\renewcommand{\theequation}{\thesection.\arabic{equation}}

\begin{flushright}
BI-TP 2014/25  
\\
\end{flushright}
\begin{centering}

\vfill

{\Large\bfseries
Order $ g ^ 2 $ susceptibilities  
   \\[2.5mm]
in the symmetric phase 
of the Standard Model
}

\vspace*{.6cm}

D.~~B\"odeker
\footnote{bodeker@physik.uni-bielefeld.de} and 
M.~Sangel
\footnote{msangel@physik.uni-bielefeld.de}

\vspace*{.6cm} 

{\em 
Fakult\"at f\"ur Physik, Universit\"at Bielefeld, 33501 Bielefeld, Germany
}

\vspace*{0.3cm}


\vspace*{1cm}
 
{\bf Abstract}

\end{centering}
 
\vspace{0.5cm}
\noindent
Susceptibilities of conserved charges such as baryon minus lepton
number enter baryogenesis computations, since they provide the relationship
between conserved charges and chemical potentials. Their next-to-leading
order corrections are of order $ g $, where $ g $ is a 
generic Standard Model coupling. They are due 
to soft Higgs boson exchange, and have been calculated recently, 
together with some order $ g ^ 2 $ corrections.
Here we compute the complete  $ g ^
2 $ contributions.  Close to the
electroweak crossover the soft Higgs contribution is of order $ g ^ 2
$, and is determined by the non-perturbative
physics at the magnetic screening scale.

\vspace{0.5cm}\noindent

 
\vspace{0.3cm}\noindent
 
\vfill \vfill
\noindent
 

 
\section{Introduction}
\label{s:intro} 

In the early Universe all charges which are violated at a rate smaller
than the Hubble expansion rate can be considered conserved. For
instance, in the minimal Standard Model (with zero neutrino masses)
baryon number $ B $ and the $ n _ {\rm f} = 3$ flavor lepton numbers $
L _ i $ are conserved below the electroweak scale, while at higher
temperatures only the differences $ X _ i \equiv B/n _ {\rm f } - L _
i $ are conserved. All equilibrium properties are determined by the
temperature $ T $ together with the values of {\it  all} conserved charges
$ Q _ i $  or
equivalently by the corresponding chemical potentials $ \mu  _ i $. 
These properties are
encoded in the grand canonical partition function
\begin{align}
  \exp (  - \Omega  /T ) 
  = { \rm tr } \exp \big [ (   \mu  _ i Q _ i - H  ) /T \big ]  
   \label{Z} 
   , 
\end{align}
where $ H $ is the Hamiltonian. 

It is rather plausible that initially the values of conserved charges
were practically zero, for example if one assumes that the Universe
underwent an early period of inflation. Since there is something
rather than nothing, some processes must have created at least the charge
that we know is non-vanishing at present, i.e., the baryon number, or
baryon asymmetry of the Universe.  Such a process, called
baryogenesis, must proceed out of thermal equilibrium. For example, in
leptogenesis \cite{fukugita} a non-vanishing value of some $ X _ i $ is generated.
Afterwards this quantity is conserved and its value determines the
equilibrium properties, such as the expectation values of baryon
number $ B $ or lepton number $ L $.

The values of the charges and thus of the chemical potentials are
usually small, so that the grand canonical potential is only needed
to lowest non-trivial order,  which is $ O ( \mu   ^ 2 ) $.%
\footnote{We assume that the charges $ Q _ i $ are odd
  under CPT. Then their expectation values vanish when $ \mu = 0 $,
  and $ \Omega $ contains no terms linear in $ \mu $.} Then the $ \mu 
$-dependence is fully determined by the second derivatives at zero $
\mu $, the so-called susceptibilities
\begin{align} 
   \chi  ^{ }_ { ij }  \equiv 
    - \frac1V 
   \left . \frac{ \partial ^ 2 {\Omega  }   }
   { \partial \mu  _ i \partial \mu  _ j }  \right | _ { { \mu }    = 0 } 
 \label{susc}   \;
 .
\end{align} 

One important use of the grand canonical potential is to determine
the relation between $ B $ or $ L $ and the $ Q _ i $.
Strictly speaking one cannot introduce a chemical potential 
for $ B + L $ in the symme\-tric phase where 
electroweak sphalerons rapidly violate $ B+L $. 
Nevertheless, one can {\em  formally } introduce a chemical potential
for  $ B + L $ as long as one computes only the expectation value
of $ B + L $ and not higher moments. The reason is that for 
the resulting partition function 
\begin{align}
  \exp ( - \Omega  ' / T ) 
  =
  { \rm tr } \exp \Big\{  
    \big[ \mu  _ { \scriptscriptstyle B + L } ( B + L)
     + \mu  _ i Q _ i  - H \big] /T \Big\}
     \label{Oprime} 
\end{align} 
one only needs the expansion 
to first order in $ \mu  _ { \scriptscriptstyle B + L } $. 
Then, 
even though $ B + L $ does not commute with $ H $,  the 
operator ordering does not matter because of the trace. 
The expectation value can then be written as
\begin{align} 
   \langle B + L \rangle 
   = - \left. \frac {  \partial \Omega ' } { 
       \partial \mu  _ { 
         B + L } } 
   \right | _ { \mu  _ { 
       B+L } =0 } 
   \label{ev} 
   \,.
\end{align} 
This relation can be used to determine  $ B + L $ and thus 
$ B $ from 
the value of $ B - L $ before the electroweak crossover, neglecting
possible effects of the non-equilibrium epoch when  the electroweak sphaleron
transitions are shut off.

Another use of the susceptibilities (\ref{susc}) has been pointed out
recently \cite{bodeker-washout} in the context of leptogenesis.  There
the asymmetry can be obtained from a set of kinetic equations.
One coefficient in these equations quantifies the amount of washout of
the asymmetry.  It was found 
that at leading
order in the right handed neutrino Yukawa couplings the washout rate
can be factorized into a product of a spectral function which contains
dynamical information, and the inverse of a matrix of susceptibilities.
The spectral function has been computed at next-to-leading order which
is $ \ord ( g ^ 2 ) $ in the Standard Model couplings $ g
$.\footnote{For our power counting we make no distinction between the
different Standard Model couplings. In this respect
we differ from  \cite{gynther}.} It 
turned out that deep in the symmetric phase the NLO corrections to the
susceptibilities already start at order $ g $.  The $ O ( g ) $
contribution computed in \cite{bodeker-washout} 
is an infrared effect caused by the exchange of a soft
Higgs boson. 
Close to the electroweak crossover the effective thermal
Higgs mass can become very small. If it becomes of the order of the
magnetic screening scale, the perturbative expansion for the
susceptibilities can be expected to break down.

In this paper we compute the complete $ O ( g ^ 2 ) $ corrections
to the susceptibilities, thereby completing the $ O ( g ^ 2 ) $ 
result for the washout rate. We obtain contributions both from 
hard ($ \sim T $) and smaller momenta, which, depending 
on the value of the thermal Higgs mass, can be  
soft ($ \sim g T $) or even smaller (`ultrasoft'). We use dimensional reduction, 
a framework which allows us to systematically treat the contributions
at the different scales and the required resummations. 

Part of the $ O ( g ^ 2 ) $ susceptibilities
have already been computed in \cite{bodeker-washout}. 
Dimensional reduction in the presence of
chemical potentials has been considered in \cite{gynther}, where the
focus was on a electroweak phase transition.  Therefore only those
terms which depend on the Higgs field were computed.

This paper is organized as follows. 
In section \ref{s:gauge}  we recall the role of gauge charges and gauge
fields in the presence of chemical potentials for global charges. 
Section \ref{s:dimred} outlines our use of dimensional reduction. The hard
Higgs contribution is obtained in section \ref{s:hard}, and the dimensionally
reduced theory is described in section \ref{s:reduced}. Depending on 
the value of the effective Higgs mass we obtain either
soft (section \ref{s:soft}) or both soft and ultrasoft
contributions (section \ref{s:ultra}). 
Finally, in section \ref{s:kappa} we illustrate our results
by computing the relation of $ B $ and $ B - L $ near the
electroweak crossover. 

\section{Chemical potentials and gauge charges}
\setcounter{equation}0
\label{s:gauge} 

We write the partition function (\ref{Z}) 
as a path integral with  imaginary time $ t = -i \tau  $, 
\begin{align}
  \exp ( -  \Omega     /T )
   =
   \int  { \cal D } \Phi  \exp \left \{  \int _ 0 ^ { 1/T }
     d \tau  
     \left [ 
        \mu  _ i Q _ i + \int d ^ 3 x 
       {\cal L } 
       \right  ] 
     \right \}
     \label{path} 
     , 
\end{align}
where $ \Phi  $ stands for all fields in our theory with the Lagrangian 
$ \cal L $.
The temporal component of the gauge
fields act as  Lagrange multipliers  which enforce Gauss' law.
We work in a finite volume and take the volume  to
infinity in the end. Then, with spatial periodic boundary conditions, the
total gauge charges vanish. These conditions are enforced by the constant 
modes of the temporal component of the gauge fields.

In the presence of chemical potentials for global charges the temporal
components of the gauge fields can
develop constant expectation values which act like chemical potentials
for the corresponding gauge charges. We will only consider the
symmetric phase of the electroweak theory, where only the weak hypercharge
gauge field $ B _ \mu $ can develop an expectation value.

It is convenient to perform the path integral (\ref{path}) 
in two steps~\cite{khlebnikov}.
First one integrates over all fields except over 
the constant
mode of $   B _ 0 $ which we denote by $ \bar{ B } _ 0 $.
We denote the result of this integration by
$ \exp ( - \widetilde{ \Omega  }   /T ) $.
In the presence of chemical potentials
$ \widetilde{ \Omega  } $ may contain terms linear in $ \bar B _ 0 $. 
The linear terms can arise when some of the global charges are 
correlated with the hypercharge. Then the  integral over $ \bar{ B } _ 0 $
\begin{align}
  \exp (  - { \Omega  }/T ) 
  = \int d \bar{ B } _ 0  
   \exp (  - \widetilde{ \Omega  } /T ) 
   \label{B0integral} 
\end{align} 
can lead to $ \mu  $-dependent contributions.

Here we are interested in small values of the conserved
charges which corresponds to small values of the chemical potentials.
Therefore we  need to keep only those terms in $ \widetilde{
  \Omega } $ which are at most quadratic in the chemical potentials.
Then (\ref{B0integral}) can be evaluated in the saddle point
approximation,
\begin{align}
  \exp (  - { \Omega  }/T ) 
  = \mbox{const} \times 
   \exp \!
   \left [ - \widetilde{  \Omega  } (  {\rm saddle \, point })   
     /T \right ] 
   \label{spa}
   .
\end{align}
Here  $ \widetilde{ \Omega  } $ is evaluated at the saddle point 
\begin{align} 
   \frac { \partial \widetilde{ \Omega  } } { \partial \bar{  B } _ 0 } = 0
   \label{saddle} 
   , 
\end{align}
and the constant in (\ref{spa}) is independent of the chemical potentials.  The
relation (\ref{saddle}) determines the expectation value  of $ \bar{ B } _ 0 $
and 
is usually referred to as `equilibrium
condition'. Note that it follows from 
the saddle point approximation to (\ref{B0integral}).

Our convention is such that the hypercharge gauge field enters
the covariant time derivative
for species $ \alpha  $ with hypercharge $ y _ \alpha  $ as follows, 
\begin{align} 
   D _ 0 
   =  \partial _ t + i y _ \alpha  g _ 1 B _ 0 
   + \cdots  
   = i ( \partial _ \tau  +  y _ \alpha  g _ 1 B _ 0 ) 
   + \cdots  
   \label{D0} 
   , 
\end{align}
where $ y _ \varphi = 1/2 $ for the Higgs field, and $ g _ 1 $ is the
weak hypercharge gauge coupling.  Note that $ B _ 0 $ is purely
imaginary.  The constant mode acts like a chemical potential $ \mu 
_ \alpha = y _ \alpha \mu _ Y $ for each species $ \alpha $ with
the `hypercharge chemical potential' 
\begin{align} 
   \mu  _ Y   \equiv      g _ 1 \bar B _ 0 
   \label{mualpha} 
   . 
\end{align}
It is, like $ \bar B _ 0 $, purely imaginary. 

\section{Dimensional reduction}
\setcounter{equation}0
\label{s:dimred} 

A useful tool for consistently treating the contributions from the
different momentum scales at high temperature is dimensional reduction
\cite{Appelquist:1981vg,farakos,braaten-scalar,kajantie}. 
The constant gauge field modes (see section \ref{s:gauge})
can also be conveniently  treated within this framework.  
Thus the computation of the grand canonical partition
function is conveniently done as follows: In a first step one
integrates out hard field modes with momenta of order $ T $. This
includes all fermion fields because in the imaginary time formalism
their (Matsubara-) frequencies cannot vanish and are always of order $
T $. The result is an effective action containing $ \widetilde{\Omega 
} _ {\rm hard } $, which aside from the zero modes is field
independent, and an effective Lagrangian $ {\cal L } _ {\rm soft } $
for a 3-dimensional field theory, and momenta of order $ g T $ or less.
In a second step one integrates over soft modes which are the zero
frequency modes with spatial momenta of order $ g T $. This yields $
\widetilde{ \Omega } _ {\rm soft } $ plus an effective Lagrangian for
the ultrasoft ($ p \ll g T $) fields $ {\cal L } _ {\rm ultrasoft } $.
When the Higgs mass in $ {\cal L } _ {\rm soft } $ is small compared
to $ g T $, there are also important contributions from an ultrasoft
spatial momentum scale smaller than $ g T $, as will be discussed
below. After these steps one obtains $ \widetilde{ \Omega } $ and from
that $ \Omega $ using~(\ref{spa}).  In this way we obtain the grand
canonical potential as a sum of three parts,
\begin{equation}
\widetilde \Omega  = \widetilde \Omega  _ {\rm hard } 
   + \widetilde \Omega  _ {\rm soft } 
   + \widetilde \Omega  _ {\rm ultrasoft } 
   .
\end{equation} 
In principle it would be possible to treat the constant mode of the
gauge fields as part of the 3-dimensional gauge field, without
introducing the notion of a gauge charge chemical potential. Then the
distinction between constant and non-constant gauge fields would have
to be made only when integrating out the soft fields. In such an
approach the mass term for $ B _ 0 $ would not only contain the Debye
mass for the soft field, but also a linear and a quadratic term in the
constant mode. This point of view was taken in \cite{khlebnikov}.
For a next-to-leading order calculation it is more convenient to
distinguish the two as in \cite{bodeker-washout}, because the masses
for the non-constant modes are only needed at order $ g ^ 2 T ^ 2 $, while $
g ^ 2 T ^ 2 \mu _ \varphi ^ 2 \sim g ^ 4 T ^ 2 \bar B _ 0 ^ 2 $.  
Furthermore, in this way we 
can easily read off the fermionic contributions to $ \widetilde{
  \Omega } $ from \cite{bodeker-washout}.

\section{Hard contributions}
\setcounter{equation}0
\label{s:hard} 

We compute $ \widetilde{ \Omega  } _ {\rm hard } $ 
in the Standard Model in  4 dimensions.
We need the terms of the Lagrangian which contain the Higgs field
$ \varphi  $, 
\begin{align} 
\mathcal{L}  _ \varphi  =  
   &   
  -\varphi^{\dagger}D^2\varphi\
  - m_{0}^{2}\varphi^{\dagger}\varphi
  - \lambda\left(\varphi^{\dagger}\varphi\right)^{2}
  \nonumber \\{} &
   - {} \left [ 
   \left(h_{e}\right)_{ab}\bar{l}_{a,L}\varphi e_{b,R}+\left(h_{u}\right)_{ab}\bar{q}_{a,L}\tilde{\varphi} u_{b,R}+\left(h_{d}\right)_{ab}\bar{q}_{a,L}\varphi d_{b,R}+
  \mbox{   h.c.} 
  \right ] 
 \label{eq:SMLagrangian}
 .
\end{align} 
We treat all particles as massless and perform a perturbative expansion
in the parameters $m_{0}^{2}$, $\lambda$, $h_i $, and $g_i$, where
$ g _ 2 $ and $ g _ 3 $ are the weak SU(2) and color SU(3) gauge couplings,
respectively. We treat all couplings as being of order $ g $, and $ m _ 0 ^ 2 
\sim g ^ 2 T ^ 2 $.
We use dimensional regularization by working in 
$ d = 3 - 2 \varepsilon  $ spatial dimensions. Then infrared
divergences coming from massless propagators va\-nish automatically.
The Higgs chemical potential (see (\ref{mualpha})) introduces the following 
additional terms:
\begin{equation}
  \delta\mathcal{L}=\mu_{\varphi}
  \left [ \varphi^{\dagger}\left(\partial_{\tau}\varphi     \right )
  -\left(\partial_{\tau}\varphi^{\dagger}\right)\varphi
 \right ] 
  +\mu_{\varphi}^{2}\varphi^{\dagger}\varphi
  +2g_{1}\mu_{\varphi}B_{0}\varphi^{\dagger}\varphi
  +2g_{2}\mu_{\varphi}\varphi^{\dagger}A_{0}\varphi
  \label{eq:muLagrangian}
  .
\end{equation}
Even though we only need an expansion up 
to order $ \mu  _ \varphi  ^ 2 $, we find it convenient
to include the quadratic term in (\ref{eq:muLagrangian}) in the Higgs propagator
and later expand the loop integrals. Note that 
there are also $ \mu  _ \varphi  $-dependent vertices 
whose effects cannot be covered by a frequency shift 
in the propagator. We will see that  the 
diagrams containing these vertices vanish
at order $\mu^{2}$ because the sum integral (\ref{J2})  is zero. 

In the calculation for the hard contributions the following 1-loop
sum-integrals $ \SumInt _ p \equiv T \sum _ { p _ 0 } \int _ { \vec p
} $ with $ \int _ { \vec p } \equiv ( 2 \pi  ) ^ { -d } \int d ^ d p $ 
appear:
\begin{align} 
  J _ 0 ( \mu  _ \varphi  ) \,
   \equiv  & \,
   \int\!\!\!\!\!\!\!\!\sum _ p \ln(- p ^ 2 ) 
     =-\frac{\pi^{2}T^{4}}{45}
     -\mu_\varphi^{2}\frac{T^{2}}{6}+\ord(\mu_\varphi^{4})
     ,
   \\
   J_ 1 ( \mu  _ \varphi  ) \equiv  & \,
   \SumInt _ {p}
   \frac { 1 } {  -p ^ 2 } 
   =\frac{T^{2}}{12}-\frac{\mu_\varphi^{2}}{8\pi^2}+\ord(\mu_\varphi^{4})
   \label{bose}
   .
\end{align}
Here and below we denote $ p ^ 2 = p _ 0 ^ 2 
-  \vec p ^ 2 $, and $ p _ 0 = i n 2 \pi  T + \mu  _ \varphi  $ with
summation over all integer $ n $.
The only  2-loop sum-integral which cannot be reduced to products of 1-loop
integrals is only needed at zero chemical potential, where it vanishes 
exactly, 
\begin{equation}
  J_{2} \equiv \SumInt_{p,q}\frac{1}{p^{2}q^{2}(p+q)^{2}} \Bigg | _ {
    \mu  =0  } =0
  \label{J2}
   .
\end{equation}
This result has been found to order $\ord (\varepsilon)$ in 
\cite{ArZh1,ArZh2}, and to all orders in  \cite{Nishimura:2012ee}.

We then obtain the following contributions to $ -\widetilde{\Omega}_{\rm hard} / V $:
The leading order is given by the 1-loop diagram
\begin{align}
\TopoVR(\AHi)= -2 
& {J}_ 0 ( \mu_{\varphi} ) 
=2\left(\frac{\pi^{2}T^{4}}{45}
+\mu_{\varphi}^{2}\frac{T^{2}}{6}+\ord(\mu_{\varphi}^{4})\right)
   \label{1loop4d} 
   .
\end{align}
There is also one 1-loop diagram with a Higgs mass insertion
\begin{align}
\TopoVRo(\AHi)= -2 
& {J}_ 1 ( \mu_{\varphi} ) 
=-2m_{0}^2\left(\frac{T^{2}}{12}-\frac{\mu_{\varphi}^{2}}{8\pi^2}+\ord(\mu_{\varphi}^{4})\right)
   \label{1loop4dmass} 
   .
\end{align}
At 2 loops we have the Higgs self interaction, 
\begin{align}
\frac{1}{2}\ToptVE(\AHi,\AHi)= & -6\lambda 
   J_ 1   ^ 2( \mu_{\varphi} ) \nonumber \\
= & -\lambda\frac{T^{2}}{2}\left(\frac{T^{2}}{12}-\frac{\mu_{\varphi}^{2}}{4\pi^{2}}\right)+\ord(\mu_{\varphi}^{4})
   \label{selfh} 
   . 
\end{align}
The results for the individual diagrams are in Feynman gauge, 
and we have checked that their sum is gauge fixing independent. 
The gauge fields carry zero chemical potentials, 
and we denote their   momenta 
by $ q $. Their interaction with
the Higgs field gives
\begin{align}
\frac{1}{2}\ToptVE(\AHi,\Aga)
   = & - \frac{d+1}{2} \left (g_{1}^{2}+3g_{2}^{2} \right )
   J_ 1 ( \mu_{\varphi} ) J_1 ( 0 ) \nonumber \\
= & -\frac{d+1}{2} \left (g_{1}^{2}+3g_{2}^{2} \right )  
\frac{T^{2}}{12}\left(\frac{T^{2}}{12}-\frac{\mu_{\varphi}^{2}}{8\pi^{2}}+\ord(\mu_{\varphi}^{4})\right), 
\end{align}
\begin{align}
  \frac{1}{2}\ToptVS(\AHi,\AHi,\Lga) 
  & =\frac{1}{4}(g_{1}^{2}+3g_{2}^{2})\SumInt_{p, q}
   \frac{(2 p+ q )^{2}} 
   { p ^{2} q ^ 2 ( p + q ) ^ 2 }\nonumber \\
 & =\frac{1}{4}(g_{1}^{2}+3g_{2}^{2})
 \left [ 4 J_ 1 ( \mu  _ \varphi  ) J_ 1 ( 0 ) 
   -J_1 ^ 2 ( \mu_{\varphi}) +\ord(\mu_{\varphi}^{4})\nonumber\right ]
   \nonumber \\
 & =\frac{1}{4}(g_{1}^{2}+3g_{2}^{2})\frac{T^{2}}{12}\left(3\frac{T^{2}}{12}-2\frac{\mu_{\varphi}^{2}}{8\pi^{2}}+\ord(\mu_{\varphi}^{4})\right).
\end{align}
Finally, the diagram 
\begin{align}
  \frac12
  \quad 
  {\pic{\AHi(15,15)(15,0,180)
      \Line(0,15)(-7,15)      
      \Text(-4 , 15)[]{$ \times $ }  
      \Line(30,15)(36,15)
      \Text(38 , 15)[]{$ \times $ }  
      \AHi(15,15)(15,180,360)%
      \Lga(30,15)(0,15)}}
  \quad
  =-\frac{1}{4} \mu  _ \varphi  ^ 2 (g_{1}^{2}+3g_{2}^{2})J_{2}
  + O ( \mu  _ \varphi  ^ 4 ) 
  = O ( \mu  _ \varphi  ^ 4 )
  .
\end{align} 
contains the 3-vertices in (\ref{eq:muLagrangian}) which are proportional
to $ \mu _ \varphi  $. Thus at second order in $ \mu  _ \varphi 
$ we can evaluate the sum-integral with zero chemical potential 
in which case it vanishes, see  (\ref{J2}).
The 2-loop contributions above contain symmetry factors 1/2 which we have
displayed as explicit prefactors of the diagrams.

All terms of the contributions to $ \widetilde{ \Omega  } $ computed in 
\cite{bodeker-washout} containing fermionic chemical potentials or
Yukawa couplings are hard.\footnote{This is easy to see since 
the integrals for diagrams with fermions 
can be written as products of 1-loop integrals.}
Therefore by combining the hard purely bosonic contributions computed
above with the ones containing 
fermions from \cite{bodeker-washout} we obtain the complete hard
contribution as
\begin{align} 
 & \hspace*{-2cm} 
 - \frac{12 }{V T^2 } \left [ 
   \widetilde \Omega  - \widetilde \Omega  (  { \mu  = 0 }  ) 
 \right ] _ {\rm hard} 
 \nonumber \\ 
 & =  
 6\, \biggl[ 1 - \frac{3}{8\pi^2} \biggl(
  \frac{g_1^2}{36} + \frac{3 g_2^2}{4} + \frac{4 g_3^2}{3}
  \biggr)\biggr]
   {\rm tr } ( \mu  _ q ^ 2 ) 
   \nonumber \\
 & +  
 3\, \biggl[ 1 - \frac{3}{8\pi^2} \biggl(
  \frac{4g_1^2}{9} + \frac{4 g_3^2}{3}
 \biggr)\biggr] 
  {\rm tr } ( \mu  _ u ^ 2 ) \nonumber  \\
 & +  
 3\, \biggl[ 1 - \frac{3}{8\pi^2} \biggl(
  \frac{g_1^2}{9} + \frac{4 g_3^2}{3}
 \biggr)\biggr] 
  {\rm tr } ( \mu  _ d ^ 2 ) \nonumber  \\
 & +  
 2\, \biggl[ 1 - \frac{3}{8\pi^2} \biggl(
  \frac{g_1^2}{4} + \frac{3 g_2^2}{4}
 \biggr)\biggr] 
  {\rm tr } ( \mu  _ \ell ^ 2 ) \nonumber \\ 
 & +  
 \;\; \biggl[ 1 - \frac{3}{8\pi^2} 
  g_1^2
 \biggr] 
  {\rm tr } ( \mu  _ e ^ 2 ) \nonumber  \\
  & +  
  4\, \biggl[ 1 
 + \frac{3}{4\pi^2} \biggl( 
  \frac{1}{2} \lambda  + 
  \frac{g_1^2 + 3 g_2^2}{8 }+
  \frac{m_{0}^2}{T^2}  
 \biggr) 
 \biggr]\, \mu_{\varphi}^2 \nonumber  \\
 & +  3 
 \biggl[ \frac{1}{4\pi^2}
    {\rm tr }( h^{ } _ u h _ u ^\dagger ) \mu_\varphi^2
 - \frac{3}{8\pi^2} {\rm tr} \Bigl(h _ u ^\dagger h^{ } _ u  \mu_{q}^2 
         + h^{ } _ u h _ u  ^\dagger \mu_{u}^2  \Bigr) 
 \biggr]^{ }
 \nonumber \\  
 & +  3 
 \biggl[ \frac{1}{4\pi^2}
 {\rm tr }( h^{ } _ d h _ d ^\dagger )  \mu_\varphi^2
 - \frac{3}{8\pi^2} 
{\rm tr} \Bigl(h _ d ^\dagger h^{ } _ d  \mu_{q}^2 
         + h^{ } _ d h _ d  ^\dagger \mu_{d}^2  \Bigr) 
 \biggr]^{ }
 \nonumber \\
 & +  
 \biggl[ \frac{1}{4\pi^2}
 {\rm tr }( h^{ } _ e h _ e ^\dagger ) \mu_\varphi^2
 - \frac{3}{8\pi^2} 
  {\rm tr} \Bigl(h _ e ^\dagger h^{ } _ e  \mu_{\ell}^2 
         + h^{ } _ e h _ e  ^\dagger \mu_{e}^2  \Bigr) 
 \biggr]^{ }
 \label{Omegahard} 
 + O(\mu^4) \;
  . 
\end{align} 
Here the $ h _ i $, are the Yukawa coupling matrices (see (\ref{eq:SMLagrangian})). 
The chemical potential matrices 
are determined by the zero mode $ \bar{ B } _ 0 $, or hypercharge chemical potential,
and by the chemical potentials
in (\ref{Z}), 
\begin{align}
  \mu  _ \alpha   = 
  y _ \alpha    \mu  _ Y 
  + \sum _ i \mu  _ i T _ { i, \alpha   } 
  \label{mumatrix} 
  .
\end{align} 
The $ T _ { i, \alpha   }  $ are the generators of the symmetry transformation
corresponding to
the charge $ Q _ i $, acting on fermion  type $ \alpha  $ with
$ \alpha  \in \{ q, u, d, \ell, e \} $. 
For example, the generator matrices of 
$ B -L $ are proportional to the unit matrix, with
$ T _ {  B-L  , q } =  T _ {  B-L , u } =  
T _ {  B-L , d } =  1/3 $ and 
$ T _ {   B-L , \ell  } 
= T _ {   B-L  , e } =  -1 $.   

\section{The dimensionally reduced theory}\setcounter{equation}0
\label{s:reduced} 

Aside from the hard contribution $ \widetilde { \Omega } _ {\rm hard }
$ the hard modes also determine the effective Lagrangian for the 
bosonic modes with zero Matsubara frequency, and with 
soft or ultrasoft momenta.  The derivation of an effective
three-dimensional theory of the Standard Model has been done in
\cite{kajantie} at zero $ \mu  $. 
At order $ g ^ 2 $ we need the following $ \mu  $-independent terms:%
\footnote{The term $ \varphi  ^\dagger A _ 0 B _ 0 \varphi  $ term does
not contribute at $ \ord ( g ^ 2 ) $.}
\begin{eqnarray}
  - \mathcal{L}_{\rm soft,\, \mu_{\varphi}=0} 
   & = & \frac{1}{4}F_{ij}F_{ij}+\frac{1}{4}W_{ij}W_{ij}\nonumber \\
   & + & \varphi^{\dagger}\vec   D^2\varphi
   +m_{3}^{2}\varphi^{\dagger}\varphi+\lambda_{3}\left(\varphi^{\dagger}\varphi\right)^{2}\nonumber \\
 & - & \frac{1}{2}(\partial_{i}B_{0})^{2}-\frac{1}{2}m_{D,1}^{2}B_{0}^{2}-\frac{1}{2}(D_{i}A_{0})^{2}-\frac{1}{2}m_{D,2}^{2}\text{Tr}\left(A_{0}^{2}\right)\nonumber \\
 & - & h_{1}\varphi^{\dagger}\varphi B_{0}^{2}-h_{2}\varphi^{\dagger}\varphi\text{Tr}\left(A_{0}^{2}\right)
   \label{Lsoft} 
   .
\end{eqnarray}
For the finite density effects we also need to include
\begin{equation}
  - \delta\mathcal{L}_{\rm soft }
  =-\mu_{\varphi}^{2}\varphi^{\dagger}\varphi-\rho_{1}\varphi^{\dagger}B_{0}\varphi-\rho_{2}\varphi^{\dagger}A_{0}\varphi.
\end{equation}
The quadratic scalar operators can be combined, yielding a
$\mu_{\varphi}$ dependent mass \cite{farakos,kajantie} 
\begin{align}
  m_{3,\mu_{\varphi}}^{2} & \equiv-\mu_{\varphi}^{2}+m_{3}^{2}\nonumber \\
  & =m_{0}^{2}-\mu_{\varphi}^{2}+T^2\left(\frac{1}{2}\lambda
    +\frac{3}{16}g_{2}^{2}
    +\frac{1}{16}g_{1}^{2}+\frac{1}{4}h_{t}^{2}  \right)
   \label{m3}
   ,
\end{align}
where $ h _ t $ is the (real) top Yukawa coupling. 
As discussed at the end of section \ref{s:dimred}, 
the Debye masses for $ A _ 0 $, $ B _ 0 $ are only 
needed at order $g^{2} T ^ 2 $ \cite{kajantie},  
\begin{eqnarray}
m_{D,1}^{2} & = & \left(\frac{N_{s}}{6}+\frac{5\nF}{9}\right)g_{1}^{2}T^{2}
   ,
   \\
m_{D,2}^{2} & = & \left(\frac{2}{3}+\frac{N_{s}}{6}+\frac{5\nF}{9}\right)g_{2}^{2}T^{2} 
   \label{mD2} 
   ,
\end{eqnarray}
where $N_{s}=1$ is the number of Higgs doublets and $\nF=3$ is
the number of families. The couplings are only needed at tree level,  
\begin{align} 
  g_{i,3}^{2} & = g_{i}^{2}T \,\, (i=1,2,3), \quad 
\lambda_{3}  = \lambda T ,\quad 
h_{1}  =  g_{1}^{2}y_{\varphi}^{2}T, \quad 
h_{2}  =  \frac{1}{4}g_{2}^{2}T
\end{align} 
and also the  new parameters in $\delta\mathcal{L}_{\rm soft}$, 
\begin{align}
\rho_{1}  =2\mu_{\varphi}y_{\varphi}g_{1}, \quad 
\rho_{2}  =2\mu_{\varphi}g_{2}
   .
\end{align}

In our calculation for the soft contributions we encounter 
the standard 1-loop integrals
\begin{align}
  I_{0}(m) & =
  \int _ { \vec k } 
  \ln(k^{2}+m^{2})=\frac{2m^{d}}{d}\frac{\Gamma(1-\frac{d}{2})}{(4\pi)^{d/2}}
  =-\frac{m^{3}}{6\pi}+\ord(\varepsilon)
  \label{I0} 
  , 
  \\
  I_{1}(m) & 
  =
  \int _ { \vec k } 
  \frac{1}{\left(k^{2}+m^{2}\right)}
  =m^{d-2}\frac{\Gamma(1-\frac{d}{2})}{(4\pi)^{d/2}}=-\frac{m}{4\pi}+\ord(\varepsilon)
  \qquad  
  .
\end{align} 
In the case $m=m_{3,\mu_{\varphi}}$ we  expand in powers of $\mu_{\varphi}^{2}$,  
\begin{align}
I_{0}(m_{3,\mu_{\varphi}}) & =-\frac{m_{3}^{3}}{6\pi} 
+ \frac{\mu_{\varphi}^{2}m_{3}}{4\pi}+\ord(\mu_{\varphi}^{4})
,
\\
I_{1}(m_{3,\mu_{\varphi}}) & =-\frac{m_{3}}{4\pi}
+\frac{\mu_{\varphi}^{2}}{8\pi m_{3}}+\ord(\mu_{\varphi}^{4})
.
\end{align}
The only 2-loop integral we need is \cite{farakos,ArZh1}
\begin{align}
I(m_{a},m_{b},m_{c})
= & \int _ { \vec k _ 1, \vec k _ 2 } 
   \frac{1}{\left ( \vec k_{1}^{2} +m_{a}^{2}\right)
     \left(\vec k_{2}^{2}+m_{b}^{2}\right)
     \left [ (\vec k_{1} + \vec k_{2})^{2}+m_{c}^{2}\right ] }\nonumber \\
= & \frac{1}{16\pi^{2}}\left [ \frac{1}{4\varepsilon}
  +\ln\left (  \frac{\bar\mu}{m_{a}+m_{b}+m_{c}}\right)+\frac{1}{2}\right ] 
   +\ord(\varepsilon).
\end{align}
where $ \bar \mu  $ is the $ \overline{ \mbox{MS} } $ scale parameter. 
In the special case $m_{a}=m_{3,\mu_{\varphi}},$
$m_{b}=m\in\{0,m_{D,1},m_{D,2}\}$ and $m_{c}=m_{3,\mu_{\varphi}}$ it
is useful to expand  in $\mu_{\varphi}^{2}$, 
\begin{align}
  I(m_{3,\mu_{\varphi}},m,m_{3,\mu_{\varphi}})
  = & \frac{1}{16\pi^{2}}
  \left [ 
    \frac{1}{4\varepsilon}
    +\ln\left(\frac{\bar\mu}{2m_{3,\mu_{\varphi}}+m}\right)+\frac{1}{2}
  \right ] 
    \\
    = & \frac{1}{16\pi^{2}}
    \left [ \frac{1}{4\varepsilon}
      +\ln   \left (  
        \frac{\bar\mu}{2m_{3}+m}
      \right)+\frac{1}{2}
      +\frac{\mu_{\varphi}^{2}}{m_{3}(2m_{3}+m)}
    \right ] 
    +\ord(\mu_{\varphi}^{4})
   \nonumber 
   .
\end{align}

\section{Soft contributions for soft 
Higgs mass
} 
\setcounter{equation}0
\label{s:soft} 

In this section we consider temperatures high enough so that $ m _ 3 ^
2 $ is of order $ ( g T ) ^ 2 $ and positive. At lower temperatures,
close to the electroweak crossover, the thermal mass squared can be
almost canceled by the negative zero temperature $ m _ 0 ^ 2$, 
making $ m ^ 2 _ 3 $ smaller
than $ O ( g ^ 2 T ^ 2 )$.  This case will be discussed in section
\ref{s:ultra}.

At 1 loop we have 
\begin{align}
\TopoVR(\AHi)= & -2T I_{0}(m_{3,\mu{\varphi}})=2T\left(\frac{m_{3}^{3}}{6\pi} - \frac{\mu_{\varphi}^{2}m_{3}}{4\pi}+\ord(\mu_{\varphi}^{4})\right)
   \label{1loop} 
   .
\end{align}
At  2 loops the Higgs self-interaction gives 
\begin{align}
\frac{1}{2} \ToptVE(\AHi,\AHi)
= & -6\lambda T^{2}
   \left [ I_{1}(m_{3,\mu_{\varphi}}) \right ] ^{2}\nonumber \\
= & -\frac{3\lambda T^{2}}{8\pi^{2}}\left(m_{3}^2-\mu_{\varphi}^2
  \right) 
  + O ( \mu  _ \varphi  ^ 4 ) 
   \label{self}
   . 
\end{align}
Note that the $ \mu _ \varphi ^ 2 $-term has the same parametric form
as the one in (\ref{selfh}).  The sum of (\ref{self}) and
(\ref{selfh}) yields the $\ord (  {\lambda} ) $ correction, that has
been computed in \cite{bodeker-washout} by a Higgs mass resummation.
The interaction between Higgs and the gauge fields gives 
\begin{align}
  \frac{1}{2}\ToptVS(\AHi,\AHi,\Lga)
  = & \frac{T ^ 2}{4}(g_{1}^2+3 g_{2}^2)
  \int_{\vec k_{1}, \vec k_{2}}
  \frac{(2 \vec k_{1}+ \vec k_{2})^{2}}{(\vec k_{1}^{2}
    +m_{3,\mu_{\varphi}}^{2}) \vec k_{2}^{2}
    [ ( \vec k_{1}+\vec k_{2})^{2}+m_{3,\mu_{\varphi}}^{2} ] }
  \nonumber \\
  = & -\frac{T ^ 2}{4}(g_{1}^2+3 g_{2}^2)
  \left \{ \left [ I_{1}(m_{3,\mu_{\varphi}}) \right ] ^{2}
    +4m_{3,\mu_{\varphi}}^{2}
    I(m_{3,\mu_{\varphi}},0,m_{3,\mu_{\varphi}})
  \right \} 
  \nonumber \\
 =& {}
 \frac{\mu_{\varphi}^2 T^{2}}{32\pi^{2}}
  \,  (g_{1}^2+3 g_{2}^2)
 \left [ \frac{1}{2\varepsilon}+\frac{1}{2}
   +2\ln\left(\frac{\bar{\mu}}{2m_{3}}\right)\right ] 
   +
   \cdots  
   \label{spatial} 
\end{align}
\begin{align}
  \frac{1}{2}\ToptVS(\AHi,\AHi,\Lqq) 
  = & -\mu_{\varphi}^{2}T^2
  \left [  
     g_{1}^2I(m_{3,\mu_{\varphi}},m_{D,1},m_{3,\mu_{\varphi}})
    +3 g_{2}^2I(m_{3,\mu_{\varphi}},m_{D,2},m_{3,\mu_{\varphi}})
    \nonumber\right ] 
  \\
=&
   -\frac{\mu_{\varphi}^{2}T^{2}}{32\pi^{2}}
   \left \{ 
      y_{\varphi}^{2} g_{1}^2
   \left [ \frac{1}{2\varepsilon}+1
     +2\ln\left(\frac{\bar{\mu}}{2m_{3}+m_{D,1}}\right)
   \right ] 
   \right .
   \nonumber\\
   & 
   \phantom {   -\frac{\mu_{\varphi}^{2}T^{2}}{32\pi^{2}} } 
   \left . 
   +  3 g_{2}^2
   \left [ 
     \frac{1}{2\varepsilon}+1+
     2\ln\left(\frac{\bar{\mu}}{2m_{3}+m_{D,2}}\right)
   \right ] 
   \right \}
   + \cdots  
   \label{A0sunrise} 
   ,
\end{align}
\begin{align}
   \frac{1}{2}\ToptVE(\AHi,\Aqq)
   & =  -\frac{1}{2} g_{1}^2T^{2}I_{1}(m_{3,\mu_{\varphi}})I_{1}(m_{D,1})
   -\frac{3}{2} g_{2}^2T^{2}
   I_{1}(m_{3,\mu_{\varphi}})I_{1}(m_{D,2})
   \nonumber \\
   &   
   = -   \frac { T^2 } { 32 \pi  ^ 2 }
   \left ( g _ 1 ^ 2 m_{3,\mu_{\varphi}}m_{D,1}
   + 3 g_{2}^2 m_{3,\mu_{\varphi}}m_{D,2}
   \right ) 
   \nonumber \\
     & 
     = 
     \frac{\mu_{\varphi}^{2}T^2}{32\pi^{2}}
     \frac 1 { 2 m _ 3 } 
     \left ( 
       g_{1}^{2} m_{D,1}      +3g_{2}^{2}     m_{D,2}
     \right ) 
     +
     \cdots  
   \label{A0eight} 
\end{align}
where we omitted terms of orders other than $ \mu  _ \varphi  ^ 2 $.
Adding up all contributions we obtain the finite result 
\begin{align}
- \frac{12 }{V T^2 }  & 
   \left [ 
     \widetilde \Omega ( \mu  )  - \widetilde \Omega  (  {  0 }  ) 
 \right ] _ {\rm soft}  
 \nonumber \\
   &=2 \mu  _ \varphi  ^ 2 
   \left \{
     - \frac{3m_{3}}{\pi T}
   + \frac{9 \lambda}{4\pi^2}
   + \frac{3}{32\pi^{2}}
   \left [ 
     g _ 1 ^ 2 C _ 1 + 3 g _ 2 ^ 2 C _ 2
     \right ] 
     \right \} 
     + O ( \mu  _ \varphi  ^ 4 ) 
   \label{Osoft} 
\end{align}
with
\begin{align} 
  C_{i}\equiv 
  \frac{m_{D,i}}{m_3}-1-4\ln \left (\frac{2 m_{3}}{2 m_{3}+m_{D,i}} \right )
  \label{Ci} 
  .
\end{align}
After integrating out the soft fields we are left with an effective theory
for the ultrasoft ones. 
For soft $ m _ 3 $ the ultrasoft sector contains
only the spatial gauge fields. At the order we are considering the 
effective Lagrangian is independent of $ \mu  _ \varphi  $, so that
this sector does not contribute to the susceptibilities, and
$ \widetilde{ \Omega  } _ { \rm ultrasoft } = 0 $.

\section{Ultrasoft Higgs mass
}
\setcounter{equation}0
\label{s:ultra} 

When $ m _ 3 ^ 2 $ in (\ref{Lsoft}) becomes small, the perturbative
expansion used in section \ref{s:soft} can break down, which 
can be seen in  (\ref{A0eight}) where $ m _ 3 $ appears in the denominator. This
term is of the same order as the soft 1-loop Higgs contribution if
$ |    m _ 3 ^ 2 | \lsim g ^ 2 T m _ { D } \sim g ^ 3 T ^ 2 
   .$ 
For such small $ m _ 3 $ it is necessary to include the Higgs field in
an effective theory for momenta $ \ll g T $, which is obtained by
integrating out the temporal components of the gauge fields.
    
First consider  $ \widetilde { \Omega } _ {\rm soft}$.
Since $ m _ 3 \ll gT $ we have to put 
$ m _ 3 =0 $ in the diagrams in section \ref{s:soft}. 
Then the only non-vanishing contribution 
comes from the diagram (\ref{A0sunrise}) with $
m _ 3 \to 0 $. The other diagrams in
section \ref{s:soft} vanish in dimensional regularization. Then
$ \widetilde{ \Omega  } _ {\rm soft } $ 
contains an infrared divergence which will cancel against an
ultraviolet divergence in $ \widetilde{ \Omega  } _ {\rm ultrasoft } 
$, leaving an order
$ g ^ 2 \ln ( 1/g ) T ^ 2 \mu _ \varphi ^ 2 $ contribution to $
\widetilde{ \Omega } $.

The effective Lagrangian for the ultrasoft fields now reads
\begin{align} 
-\mathcal{L}_{\rm ultrasoft} 
    =  \frac{1}{4}F_{ij}F_{ij}+\frac{1}{4}W_{ij}W_{ij}
  -  \varphi^{\dagger}\vec D^2\varphi
  +\mbar_{3, \mu  _ \varphi  }^{2}\varphi^{\dagger}\varphi
  +\bar{\lambda}_{3}\left(\varphi^{\dagger}\varphi\right)^{2}
   \label{Lultra} 
\end{align} 
with the parameters \cite{kajantie}
\begin{eqnarray} 
\mbar_{3}^{2} & = & m_{3}^{2}
-\frac{1}{4\pi}\left(3h_{2}m_{D,2}+y_{\varphi}h_{1}m_{D,1}\right)
   \label{mbar} 
\\
\bar{\lambda}_{3} & = & \lambda_{3}.
\end{eqnarray}
The negative $\ord ( g^{3}T ^ 2) $ contribution
to $\mbar_{3} ^ 2$ results from integrating out 
the temporal components of the gauge fields. It 
leads to interesting effects depending on {\em how} soft $\mbar_{3}$
is. 

Here we have to distinguish several cases. Consider first 
$ \overline{ m } _ 3 ^ 2 \sim g ^ 3 T ^ 2$ 
and positive. 
Then we are still in the symmetric phase. The loop expansion 
parameter is now $ g ^{ 1/2 } $. The next-to-leading order (NLO) 
starts only at $  O ( g ^{ 3/2 } ) $ coming from the 1-loop diagram (\ref{1loop}), 
and the 2-loop diagrams (\ref{self}) and (\ref{spatial}) 
contribute at order $ g ^ 2 $. Combining this with the soft 
contribution we find
\begin{align}
- \frac{12 }{V T^2 }  & 
  \left [ 
   \widetilde \Omega ( \mu  )  - \widetilde \Omega  (  {  0 }  ) 
 \right ] _ {\rm soft+ultrasoft}  
\nonumber \\
   &=2 \mu  _ \varphi  ^ 2 
   \left [ 
     - \frac{3\bar{m}_{3}}{\pi T}
   + \frac{9 \lambda}{4\pi^2}
   + \frac{3}{32\pi^{2}}
   \left ( 
     g _ 1 ^ 2 \bar{C} _ 1 + 3 g _ 2 ^ 2 \bar{C} _ 2
     \right ) 
     \right ] 
   \label{Oultrasoft} 
   .
\end{align}
with
\begin{align} 
  \bar{C}_{i}\equiv 
  -1-4\ln \left (\frac{2 \overline{m}_{3}}{m_{D,i}} \right )
  \label{barCi} 
  .
\end{align}
Note that in this expression we have parametrically 
$ \ln ( m _ {{  D }  i} /  \overline{ m } _ 3 ) \sim \ln ( 1/g ) $.

There is another way to obtain (\ref{Oultrasoft}).
Since we are only  interested in the $ O ( \mu  _ \varphi  ^ 2 ) $ terms
we can expand  the path integral
\begin{align} 
   \exp ( - \widetilde{ \Omega  } _ {\rm ultrasoft } /T ) 
   = 
   \int { \cal D}\Phi _ {\rm ultrasoft} 
   \exp \left \{ 
            \int d ^ 3 x { \cal L } _ {\rm ultrasoft } 
       \right \} 
\end{align} 
to second order in $ \mu  _ \varphi  $. In (\ref{Lultra})  $ \mu  _ \varphi  $
only appears in the effective Higgs mass so that
\begin{align}
  \left [ 
    \widetilde{ \Omega  } ( \mu  ) - \widetilde{ \Omega  } ( 0 ) 
    \right ] _ {\rm ultrasoft }
    = 
    - V T \mu  _ \varphi  ^ 2 \left \langle
                              \varphi  ^\dagger \varphi  
                           \right \rangle 
    + O ( \mu  _ \varphi  ^ 4 ) 
    .
    \label{magnetic} 
\end{align} 
The expectation value of $ \varphi  ^\dagger \varphi  $ has been 
extracted
from the 2-loop effective potential 
\cite{farakos, kajantie, nonrel},
\begin{equation}
  \langle \varphi  ^\dagger \varphi  \rangle 
  _ {\rm 2-loop}
  =
  {} -\frac{\overline{  m} _ 3T}{2\pi}
   +\frac{T^2}{16\pi^2}
   \left\{
   6\lambda+
   \left(
     g_{1}^2+3g_{2}^2
   \right)
   \left [ 
     \frac{1}{4\varepsilon}
     +\ln \left (
       \frac{\bar\mu}{2 \overline{m} _ 3 } 
         \right ) 
       +\frac{1}{4}
     \right ] \right\}
    \label{phiPert} 
    ,
\end{equation}
which again leads to (\ref{Oultrasoft}).

However, (\ref{magnetic}) is also valid when $ \overline{ m } _ 3 $ becomes
as small as the magnetic screening scale $ g  ^ 2 T $ of the electroweak
theory. In this case the only momentum scale left is  $
g  ^ 2 T $. In a non-abelian gauge theory the physics at this scale is 
non-perturbative, and the loop expansion can no longer be applied,
which is the so called Linde problem \cite{Linde:1980}.
Nevertheless, the expansion in $ g $ (modulo logarithms) still exists, only 
the numerical coefficients in the series cannot be computed by summing diagrams.

Since the 3-dimensional fields have mass dimension 1/2, and since the only 
mass scale in the ultrasoft theory is $ g  ^ 2 T $, we have 
$ \langle \varphi  ^\dagger \varphi  \rangle \sim g  ^ 2 T $. 
Thus the ultrasoft fields contribute to $ \widetilde{ \Omega  } $ at order
$ g ^ 2 $. 
A reliable determination of $\langle \varphi ^\dagger \varphi \rangle$
can only be done by lattice simulation of the 3-dimensional gauge plus
Higgs system. A recent lattice study with $m_H=(125-126)$ GeV for a
SU(2)+Higgs theory can be found in \cite{D'Onofrio:2014kta}.  An older
but more comprehensive study of the SU(2) theory can be found in
\cite{Kajantie:1995kf} and a study including the U(1) gauge fields has
been performed in \cite{Kajantie:1996qd}. Near the electroweak crossover 
$ \langle \varphi ^\dagger \varphi \rangle$ turned out to be a rather smooth function
of the temperature.

Finally, for negative $ \mbar_{3}^{2} $ the Higgs field develops an expectation
value, which in presence of chemical potentials for global charges
also leads to a non-zero expectation value of 
the temporal component of the $ SU(2) $-gauge field \cite{khlebnikov}.
We have not studied this case.

\section{Relation between $B$ and $B-L$}
\setcounter{equation}0
\label{s:kappa} 

To illustrate the use of our results for $\widetilde\Omega$ we compute
the relation between the baryon number $B$ and $B-L$ in the symmetric
phase, which was done in \cite{khlebnikov} at leading order and
non-zero Higgs expectation value.  First we enforce the saddle point
condition (\ref{saddle}) to determine $ \Omega '$ as defined in
(\ref{Oprime}), thereby eliminating $\bar{B}_0$.  Then using
(\ref{ev}) and similarly for $ \langle B - L \rangle $ we express
the chemical potentials in terms of $ B \equiv \langle B \rangle $ and
$ L \equiv \langle L \rangle $ which yields a relation
\begin{align} 
  B=\kappa (B-L)
  \label{B} 
  .
\end{align} 
For $ m _ 3 $ of order $ g T $ we obtain using 
(\ref{Osoft})
\begin{align}
 \kappa=&\frac{4(2\nF+N _ s)}{22\nF+13N _ s}
 +\frac{m_{3}}{\pi  T}\frac{24 \nF N _ s }{(22 \nF + 13 N _ s)^2}\nonumber\\
 &+ \frac{g_{1}^2}{16\pi^2}\frac{236 \nF^2 - (12 C_{1}-212 ) \nF N _ s 
   + 75 N _ s^2}{(22 \nF+13 N _ s)^2}\nonumber\\
 &+ \frac{g_{2}^2}{16\pi^2}\frac{9 (12 \nF^2 - 4 (C_{2}-1 ) \nF N _ s + 3 N _ s^2)}{(22 \nF + 13 N _ s)^2}\nonumber\\
 &- \frac{g_{3}^2}{16\pi^2}\frac{96 (8 \nF^2 + 11 \nF N _ s + 3 N _ s^2)}{(22 \nF + 13 N _ s)^2}\nonumber\\
 &+\frac{h_{t}^2}{16\pi^2}
 \frac{6 (6 \nF^2 - 41 \nF N _ s - 18 N _ s^2)}{(22 \nF + 13 N _ s)^2}\nonumber\\
 &-\frac{\lambda}{16\pi^2}\frac{384 \nF N _ s}{(22 \nF + 13 N _ s)^2}\nonumber\\
 &-\frac{m_{0}^2}{(\pi T)^2}\frac{12\nF N_s}{(22\nF +13N_s)^2},
 \label{kappa}
\end{align}
with the same definitions as in (\ref{mD2}) and (\ref{Ci}). 
When $\bar
m_{3}^2\sim g^3 T^2 $ the result for $ \kappa  $ can be obtained from 
(\ref{kappa}) by replacing $m_{3}$ by
$\overline{ m}_{3}$ and $C_{i}$ by $\bar{C}_{i}$ defined in (\ref{mbar}) and
(\ref{barCi}). 
\begin{figure}[t]
	\centering
    \input{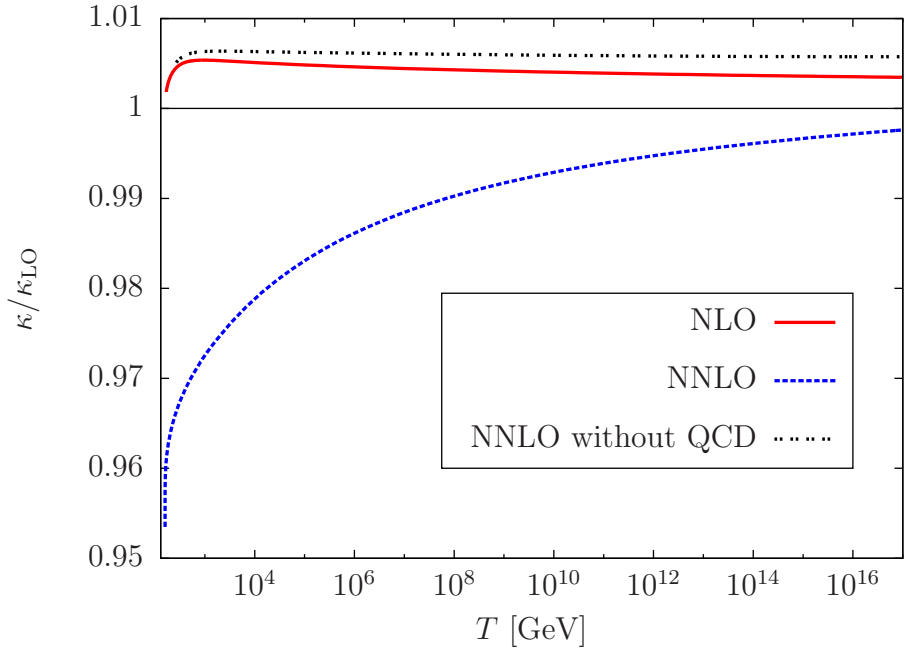}
    \caption{Size of the radiative corrections to $ \kappa  $ defined in (\ref{B}) 
      relative to the leading order result
      with $m_H=126$ GeV. The electroweak corrections are rather small, 
      and the perturbation series is well behaved.
      The complete NNLO is 
      dominated by the QCD corrections except at the highest temperatures. 
    }
	\label{fig2}
\end{figure}

\begin{figure}[t]
	\centering
   \input{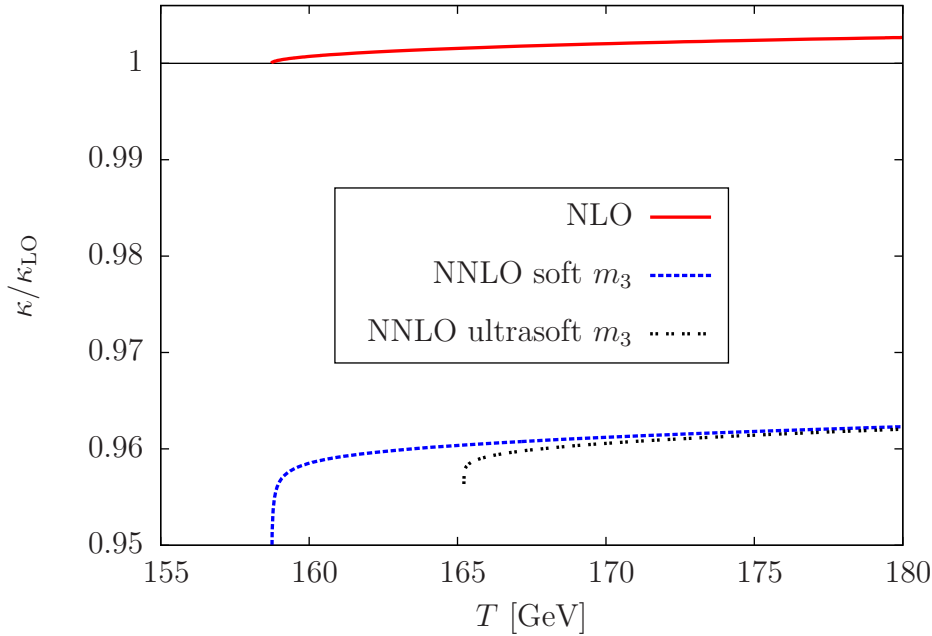}
	\caption{The ratio of $B$ and $B-L$ at low temperatures with $m_H=126$
          GeV. Shown are the LO, NLO and the NNLO result with soft and
          ultrasoft effective Higgs masses.  
           }
	\label{fig1}
\end{figure}

The size of the corrections to  $ \kappa  $ are shown in figure \ref{fig2} over
a wide range of temperatures. The next-to-leading (NLO)
corrections are entirely due to the Higgs, and they are quite small. The 
next-to-next-to-leading order (NNLO) is significantly larger. This is caused
by the relatively large QCD corrections. When the QCD corrections are
left out, the remaining NNLO corrections are even smaller than the NLO, 
indicating that the perturbation series is well behaved. We also find that
the NNLO Higgs correction has about the same size as the electroweak
corrections coming from other chemical potentials.

Figure \ref{fig1} shows a closer look at the most interesting region
near the electroweak crossover at $T\sim 160$~GeV. When $ m _ 3 $
is treated as soft, the NNLO corrections diverge like $ 1/m _ 3 $ when 
$ m _ 3 $ approaches zero. The perturbation series should be improved at small
$ m _ 3 $ by assuming $ \overline{ m } _ 3 \sim g ^ { 3/2 } T $ 
and using (\ref{Oultrasoft}).  It then diverges logarithmically
when $ \overline{ m }
_ 3 $ vanishes. Clearly, the loop expansion breaks down here. However, 
since $ \langle \varphi  ^\dagger \varphi  \rangle $ is rather smooth
when computed non-perturbatively 
on the lattice, we expect that the result for $ \kappa  $
using (\ref{magnetic}) with the non-perturbative 
$ \langle \varphi  ^\dagger \varphi \rangle  $ 
\cite{D'Onofrio:2014kta,Kajantie:1995kf,Kajantie:1996qd} 
should be rather smooth as well. 

\section{Conclusions}
\setcounter{equation}0

We have computed  the $ O ( g ^ 2 ) $ 
Higgs contribution to the  susceptibilities
in the symmetric phase of the Standard Model, thus completing the
$ O ( g ^ 2 ) $ calculation  of \cite{bodeker-washout}. 
Close to the electroweak crossover the loop expansion breaks down, 
and the  infrared Higgs contributions 
are determined by the non-perturbative electroweak magnetic 
screening scale $ g  ^ 2 T $. Nevertheless, the corrections are 
parametrically of order $ g ^ 2 $. We have obtained a relation which
can be used to determine its coefficient by a 
lattice simulation
of the 3-dimensional gauge field plus Higgs theory. 
We have
applied our result 
to compute the relation of $ B $ and $ B - L $. The corrections
are small in the regime where perturbation theory is valid. Our results indicate
that this holds even when perturbation theory breaks down. We find that
the QCD corrections dominate except at the highest temperatures, and
that the corrections are below 5\%. 

For leptogenesis our result completes the $ O ( g ^ 2 ) $ computation 
of the washout rate \cite{bodeker-washout}. Now two out of  three 
rates\footnote{The radiative corrections to the production rate are known
both in the non-relativistic \cite{nonrel,salvio}  and relativistic regime
\cite{relat}.}   
entering leptogenesis computations have been obtained at this order, 
the only missing piece being the $ CP $-asymmetry.

%
\section*{Acknowledgements}
We would like to thank H.~Nishimura, M.~Laine, and S.~Sharma
for useful discussions and suggestions.

%
\appendix 
\renewcommand{\theequation}{\thesection.\arabic{equation}}


\end{document}